\documentclass[pdflatex,sn-aps,Numbered]{sn-jnl}

\usepackage{graphicx}%
\usepackage{multirow}%
\usepackage{amsmath,amssymb,amsfonts}%
\usepackage{amsthm}%
\usepackage{mathrsfs}%
\usepackage[title]{appendix}%
\usepackage{xcolor}%
\usepackage{textcomp}%
\usepackage{manyfoot}%
\usepackage{booktabs}%
\usepackage{algorithm}%
\usepackage{algorithmicx}%
\usepackage{algpseudocode}%
\usepackage{listings}%

\theoremstyle{thmstyleone}%
\theoremstyle{thmstyletwo}%

\theoremstyle{thmstylethree}%

\begin{document}

\title{Bridging atomic and mesoscopic length scales with Replica Scanning Tunneling Microscopy: Visualizing the atomic lattice of UTe$_2$ and the atomic scale superconducting gap modulations of FeSe close to micron length scales}


\author[1]{\fnm{Miguel} \sur{\'Agueda Velasco}}
\author[1,8]{\fnm{Jose D.} \sur{Berm\'udez-P\'erez}}
\author[1]{\fnm{Pablo} \sur{Garc\'ia Talavera}}
\author[1]{\fnm{Raquel} \sur{S\'anchez-Barquilla}}
\author[1]{\fnm{Jose Antonio} \sur{Moreno}}
\author[2,3,4]{\fnm{Juan}\sur{Schmidt}}
\author[2,3]{\fnm{Sergey L.}\sur{Bud'ko}}
\author[2,3]{\fnm{Paul C.} \sur{Canfield}}
\author[5]{\fnm{Georg} \sur{Knebel}}
\author[5]{\fnm{Midori} \sur{Amano Patino}}
\author[5]{\fnm{Gerard} \sur{Lapertot}}
\author[5]{\fnm{Jacques} \sur{Flouquet}}
\author[5]{\fnm{Jean Pascal} \sur{Brison}}
\author[6]{\fnm{Dai} \sur{Aoki}}
\author[7]{\fnm{Paula} \sur{Giraldo-Gallo}}
\author[8]{\fnm{Jose Augusto} \sur{Galvis}}
\author[1]{\fnm{Isabel} \sur{Guillam\'on}}
\author[1]{\fnm{Edwin} \sur{Herrera Vasco}}
\author*[1]{\fnm{Hermann} \sur{Suderow}}\email{hermann.suderow@uam.es}

\affil[1]{\orgdiv{Laboratorio de Bajas Temperaturas y Altos Campos Magn\'eticos, Departamento de F\'isica de la Materia Condensada, Instituto Nicol\'as Cabrera and Condensed Matter Physics Center (IFIMAC), Unidad Asociada UAM-CSIC}, \orgname{Universidad Aut\'onoma de Madrid}, \orgaddress{ \city{Madrid}, \postcode{E-28049}, \country{Spain}}}

\affil[2]{\orgdiv{Department of Physics and Astronomy},
\orgname{Iowa State University},
\orgaddress{\city{Iowa}, \postcode{50011}, \country{USA}}}

\affil[3]{\orgdiv{Ames National Laboratory},
\orgname{Iowa State University},
\orgaddress{\city{Iowa}, \postcode{50011}, \country{USA}}}

\affil[4]{\orgdiv{Departamento de Física, FCEyN},
\orgname{Universidad de Buenos Aires},
\orgaddress{\city{Buenos Aires}, \postcode{1428}, \country{Argentina}}}


\affil[5]{\orgdiv{Univ. Grenoble Alpes, CEA, Grenoble INP, IRIG, PHELIQS},
\orgaddress{38000 Grenoble},
\country{France}}

\affil[6]{\orgdiv{Institute for Materials Research, Tohoku University},
\orgaddress{Ibaraki},
\postcode{311-1313},
\country{Japan}}

\affil[7]{\orgdiv{Department of Physics},
\orgaddress{Universidad de Los Andes, Bogot\'a},
\postcode{111711},
\country{Colombia}}

\affil[8]{\orgdiv{School of Sciences and Engineering},
\orgaddress{Universidad del Rosario, Bogot\'a},
\postcode{111711},
\country{Colombia}}


\abstract{Scanning Tunneling Microscopy (STM) is a cornerstone technique for visualizing the electronic density of states with atomic resolution (typically below 0.1 nm). While the field of view of most STM setups extends up to a few microns, obtaining atomic resolution over these large areas is often impractical and excessively time-consuming. This is due to the need to acquire maps with a point number reaching $10^7$ or more with a full current or conductance vs voltage curve at each point. The standard procedure is to make large scale maps and then select small regions to zoom-in for high-resolution atomic scale analysis. However, this approach fails to address a question which is often critical: Does a specific atomic-scale modulation of the electronic density of states persist over much larger, mesoscopic length scales? Here we present a new method: Replica STM (R-STM), that overcomes this limitation, allowing the study of atomic-scale phenomena up to micron length scales. We obtained new large-area STM tunneling conductance maps in UTe$_2$ and FeSe, spanning areas over 200 nm in size. In these large scale maps we discovered periodic signals with wavelengths significantly exceeding interatomic distances. We show that these large-wavelength periodic signals are replicas of the underlying atomic-scale density of states modulations. R-STM leverages these replica signals to efficiently track atomic-scale features over large areas. We discuss the influence of phase slips, disorder and defects in the replicas. Our results suggest that atomic scale modulations of the superconducting density of states could persist over large length scales in FeSe. R-STM provides a powerful and practical new capability for STM to compare atomic scale with micrometer scale phenomena. The proof of principle of R-STM can be extended to any other scanning probe microscopy experiment where a periodic signal is traced as a function of position.}


\maketitle

\section{Introduction}\label{sec1}

The invention of the Scanning Tunneling Microscope (STM) opened an unprecedented path to the visualization of matter at atomic scale\,\cite{RevModPhys.59.615}. Thanks to the capability to measure the tunneling conductance with atomic precision as a function of the voltage, it also provided a completely new method to obtain the electronic density of states of materials. In addition to the atomic precision, the tunnel junction was made through vacuum, which considerably simplified previous approaches to tunneling spectroscopy using an insulating barrier\,\cite{Wolf2012}. The electronic density of states is directly linked to the features of the ground state at low temperatures. Examples are the gap opened around the Fermi level in superconductors, which was measured shortly after the invention of the STM\,\cite{PhysRevLett.54.2433,10.1063/1.97795,Stern1987}, and the observation of the superconducting vortex lattice\,\cite{PhysRevLett.62.214,Suderow_2014,RevModPhys.79.353}. Other relevant features studied with STM are those associated with charge density waves\,\cite{PhysRevB.39.5496}. More recently, spatially modulated superconductivity, where the superconducting gap function $\Delta_{\vec{\mathbf{q}}}(\vec{\mathbf{r}})$ varies periodically in space with a characteristic momentum $\vec{\mathbf{q}}$ which is generally larger but close to atomic lattice wavevectors, has been also addressed\,\cite{AgterbergDavis2020,Du2020,Wang2021,Hamidian2016,Chen2022,Zhao2023, Liu2023,Chen2021,Deng2024,Gao2024,Sumita2025,Kong2025,Gu2023,PhysRevB.77.134505}. These and other studies \cite{Dobrovolskiy2026} raise a relevant question which is often difficult to address: Up to which length scale remain atomic size modulations coherent (i.e.\,presenting the same wavevector)?

The difficulties associated with addressing this question are mostly technical. Atomically flat terraces several $\mu$m large are often found in many materials. However, mapping the density of states with atomic precision requires a point density which is difficult to obtain and implies very long term measurements. For example, a usual cryogenic STM might have a bandwidth of about 10 kHz\,\cite{Stroscio2010,FranLHA2021,Marta22T2021}. A full conductance vs voltage curve with, say, 200 points, often requires at least 40 ms. Mapping an area 500 nm in size with atomic precision requires, say, $(5000)^2$ full conductance vs voltage curves \cite{science.1072640,science.1066974}. This requires over 11 days to complete. Often, the tip suffers an instability during the acquisition and maps need to be repeated. The process can eventually be optimized by measuring with fewer points or increasing the bandwidth. However, the bandwidth is limited by the need of a high energy resolution at low temperatures and the resonance frequency of the elements with which the microscope is built, as for example the piezotube\,\cite{Stroscio2010,FranLHA2021,Marta22T2021}.

One possible solution is adaptive sparse sampling, where a random matrix of points is used in an iterative procedure until a certain signal is recovered, instead of a full scanning grid\,\cite{nakanishi2016,Oppliger2020,Zengin2021,Oppliger2022}. As we will discuss below, phenomena with well-defined wavevectors in reciprocal space can be more efficiently studied with the technique proposed here, which is based on the periodicity of both, acquisition technique and signal from the sample. Completely different approaches, such as pump-probe techniques, address ultra high frequency phenomena\,\cite{ESR_STM2021,Loth2010}, which are somewhat different from the direct measurement of the electronic density of states.

\begin{figure}[!htb]
    \centering
    \includegraphics[width=1\linewidth]{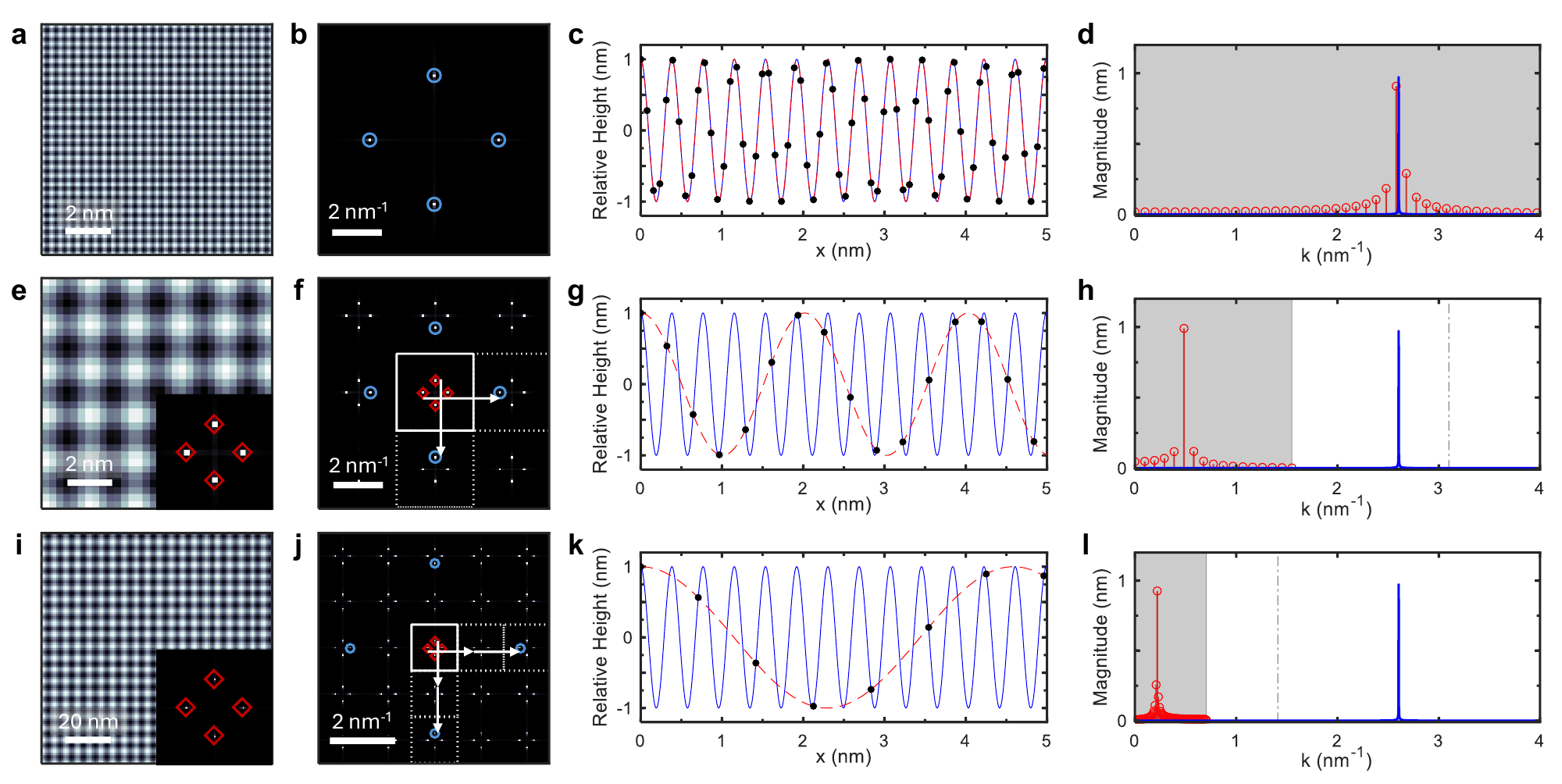}
    \caption{{\bf Simulated STM topography images.} A sinusoidal square lattice with spacing of $0.384\,\mathrm{nm}$ is sampled on three different sampling grids. (a–d) Fine sampling of a 10 nm $\times$ 10 nm area with 128$\times$128 points. In this image, $k_{S,0}= 12.7\, \frac{1}{nm}\gg 2  k_{signal}=2 \times 2.6\, \frac{1}{nm}$. In (a) we show the sampled image. In (b) we show its Fourier transform. We find Bragg peaks at $k_{signal}$ (blue circles). In (c) we show a line profile along the \(x\) axis. In blue we show the periodic signal. With black dots we show the points sampled by the image (a). The red line shows the wavefunction obtained from the black points. (d) Fourier transform of the black points (red stems and circles) and of the signal (blue line) of (c). (e–h) We now take the same square lattice and provide images using only 32$\times$32 points. Now $k_{S,1}= 3.1\, \frac{1}{nm} < 2 k_{signal}=2 \times 2.6\, \frac{1}{nm}$. In (e) we show the resulting pattern and in the inset of (e) its Fourier transform. In (f) we enlarge the Fourier space of (e) by copying the Fourier transform (e) eight times to reach the size in reciprocal space of (b). We mark with a blue circle the Bragg peaks of (b) with the original periodicity $k_{signal}$. We mark with red squares the Bragg peaks of the pattern shown in (e), $k_{replica,1}$. Arrows connect red with blue peaks and are located, for clarity, slightly shifted from Bragg peaks. In (g) we show a profile along the $x$-axis. With a blue line we show the original periodicity. Black points show the points in (e) and the red line joins the latter black points. In (h) we show the Fourier transform of (g). Blue line is the Fourier transform of the original signal and red circles and stems of the replica. (i–l) Here we enlarge field of view to 90 nm lateral size and take 128$\times$128 points. Now $k_{S,2}= 1.41\, \frac{1}{nm} < 2k_{signal}=2 \times 2.6\, \frac{1}{nm}$. The Fourier transform is shown in the lower right inset of (i). In (j) we copy the Fourier transform of the inset of (i) twenty four times to obtain a similarly sized reciprocal space as in (b). The Bragg peaks corresponding to the replica, $k_{replica,2}$ are marked by red squares. The original Bragg peaks (b) are marked as blue circles. In (k) we show as black points the points building up the image (i) along a line. The red dashed line shows a periodic function connecting black points. In blue we show the generating periodic signal. In (l) we show the Fourier transform of (k). We now have $k_{replica,2}=|k_{signal}-nk_{S,2}|$ with $n=2$. We show as a grey shadow the Nyquist band, given by $\frac{k_S}{2}$, as well as the sampling frequency, shown as a grey dashed line. In the following figures we use replica Bragg peaks (red) for the replica signal in reciprocal space and atomic Bragg peaks (blue) for the atomic scale modulation.}
    \label{Example}
\end{figure}

On the other hand, it is well known that periodic signals present replicas at wavenumbers larger than the wavenumber of the original periodicity, if the periodic signal is sampled with another fixed periodic ramping system. Nyquist theorem states that, in order to accurately map a periodic signal with a wavenumber $k_{signal}$, the sampling rate $k_S$ must be at least twice the wavenumber of the target signal $k_S \geq 2k_{signal}$. In our case, $k_{signal}$ can be for example $k_{signal}=1/a$ where $a$ is the interatomic distance. By contrast, $k_S$ is given by $k_S=1/d$ where $d$ is the distance between points in real space, or the number of points along a line of the scan divided by the lateral size of the map. Essentially this means at least two points per atom should be taken to get atomic resolution. With a smaller point density, a signal at a smaller wavenumber, $k_{replica}$, is found. $k_{replica}$ is uniquely connected to $k_S$, because $k_{replica}=|k_{signal}-nk_S|$, where $n$ is an integer given by rounding the ratio of $k_{signal}$ with $k_S$ to the nearest integer, $n=\left\lfloor \frac{k_{signal}}{k_S} \right\rceil$. We note that the sign of the wavenumbers can be ignored, as the Fourier transform of a real periodic signal satisfies Hermitian symmetry and is thus symmetric by inversion---a peak at $\mathbf{k}$ occurs also at $-\mathbf{k}$. We can illustrate the shape in real space of replica signals and their signatures in reciprocal space by an example, see Fig.\,\ref{Example} (here for a two-dimensional image).

In Fig.\,\ref{Example}(a) we show a square lattice with spacing $\lambda_{signal}=0.384\,\mathrm{nm}$. In Fig.\,\ref{Example}(b) we show its Fourier transform presenting peaks at $k_{signal}=\frac{1}{\lambda_{signal}}=2.6$ nm$^{-1}$. The square lattice has been obtained by plotting a periodic wavefunction on top of a matrix of points. The periodic wave function along the x-axis is shown in Fig.\,\ref{Example}(c) by the blue line. A profile along x of the point matrix (data plotted in Fig.\,\ref{Example}(a)) is shown as black points. The red line shows the wavefunction described by the black points. In Fig.\,\ref{Example}(d) we compare the Fourier transform of the original signal (blue in Fig.\,\ref{Example}(c)) with the one obtained from the profile of the image in Fig.\,\ref{Example}(c). The use of points in red shows that sampling is discrete. The map in Fig.\,\ref{Example}(a) contains 128$\times$128 points and is 10\,nm$\times$10\,nm large. Thus, the distance in real space between black points in Fig.\,\ref{Example}(c) is of about 0.079\,nm. There are about five points per period (black dots per period of the blue curve in Fig.\,\ref{Example}(c)). More exactly $k_{S,0}=12.7$\,nm$^{-1}$ and $k_{S,0}\gg 2 k_{signal}$. Most important is that the Fourier transform of the blue (original) and red (sampled) data presents a peak at the same value, at $k_{signal}$.

We can now reduce the number of points in the map, maintaining the same wave function to generate the map. As we show in Fig.\,\ref{Example}(e) (same lateral size but just 32 points), the result is also a periodic pattern. In this case, Nyquist theorem is not met. Here we have that $k_{S,1}=3.1$\,nm$^{-1}$, so that $k_{S,1}<2 k_{signal}$. In the Fourier transform of Fig.\,\ref{Example}(e), shown in Fig.\,\ref{Example}(f) we find four peaks from the new pattern. These are located at a value $k_{replica,1}<k_{signal}$. We can understand this by looking at the original wave function, shown by a blue line in Fig.\,\ref{Example}(g), and comparing it with the wavefunction obtained from the black points in Fig.\,\ref{Example}(g), shown as a dashed red line. We see that the red line also follows a periodic pattern, but with a much larger wavelength. The latter provides the pattern of Fig.\,\ref{Example}(e) and the peaks in its Fourier transform (Fig.\,\ref{Example}(f)). The Fourier transform now presents a peak at $k_{replica,1} = |k_{signal} - n k_{S,1}|$. In this case $n=1$ and $k_{replica,1} = |2.6-3.1|=0.5$ nm$^{-1}$. Using the relation between $k_{replica,1}$ and $k_{signal}$, we can identify $k_{replica,1}$ as a consequence of the modulation at $k_{signal}$ in a large under-sampled map. Importantly, if we observe a modulation at $k_{replica}$ in a large size map, we can deduce that the modulation at $k_{signal}$ is present over the whole map (provided that the latter is known a-priori).

We can also increase the field of view to 90\,nm$\times$90\,nm maintaining 128$\times$128 points. Now $k_{S,2}=1.41$\,nm$^{-1}$ and again $k_{S,2} < 2k_{signal}$. Although the map is heavily under sampled, there is a periodic pattern, as shown in Fig.\,\ref{Example}(i). The corresponding Fourier transform (Fig.\,\ref{Example}(j)) shows four peaks and the one-dimensional modulation shows a modulation at a very small $k_{replica,2}$ (Fig.\,\ref{Example}(k,l)). Now, $n=2$ and  $k_{replica,2} = |2.6-2.82|=0.22$ nm$^{-1}$.

We note that the undersampled signals present a modulation amplitude which is similar to the one of the original signal, as shown by the black points in Fig.\,\ref{Example}(c,g,k) and the size of the red peak in Fig.\,\ref{Example}(d,h,l).

We discuss the role of domains, random noise and defects in the Appendix. Domains produce Bragg peak broadening along the direction of domains. Random noise produces background in Fourier space and reduces signal to noise ratio. Defects lead to stripes in reciprocal space related to the distance between defects and peak broadening along the main axis in reciprocal space.

We can now concretely address the presence of replica Bragg peaks in STM images.
In general, as we shall see below, we can apply the same principle to any signal on an arbitrary direction in two dimensions by translating the replica Bragg peak along the $x$ and $y$ axis by an integer multiple of the sampling wavenumber on each direction, $n_{x,y} \mathbf{k^{x,y}_{S}}$ and find the atomic scale Bragg peaks using the relationship between $k_{replica}$ and $k_{signal}$. We represent these translations as white arrows in Fig.\,\ref{Example}(f,j). 

\section{STM Methods}\label{sec2}
To perform the topographic STM and the tunneling conductance maps we use an STM setup described in Refs.\,\cite{FranLHA2021,Marta22T2021,SuderowRSI2011,10.1063/1.4905531}. We mostly focus here on data taken from 1\,K to 5\,K. We use tips of PtIr, which we cut mechanically and sharpen in-situ by repeated indentation on a gold sample\cite{Rodrigo04}. For the topographic images, taken in the constant current mode, we use square grids, generally 256$\times$ 256 large. The tunneling conductance maps are obtained by acquiring full current vs voltage curves, each one of 256 points, at each point of a square grid of a similar size and subsequently taking the numerical derivative of the current vs voltage. We cleave samples in-situ using the method described in \cite{Marta22T2021,SuderowRSI2011}. In UTe$_2$ we obtain flat surfaces which show the (011) crystallographic direction of the orthorhombic crystal structure, as in previous work\,\cite{talaveraCDW2025,Aishwarya2023,Gu2023,Aishwarya2024,LaFleur2024,science.adk7219}. Single crystals of UTe$_2$ were grown using the chemical vapor transport method, as described in Refs.\,\cite{Aoki2022,doi10.1126science.aav8645}. The starting materials, U and Te were sealed under vacuum in a quartz ampoule with an atomic ratio of 1:1.5 together with iodine as transport agent. The ampoule was gradually heated in a horizontal furnace, and the temperature gradient 1050/990 $^{\circ}$C was applied and maintained for 10-14 days. Single crystals were obtained at the lower temperature side. FeSe crystals were obtained by vapour growth, as described in Ref.\,\cite{PhysRevB.94.024526}.

\section{Results}\label{sec3}

\subsection{Atomic lattice of UTe\texorpdfstring{$_2$}{\_2}}
\begin{figure}[!htbp]
    \centering
    \includegraphics[width=1\linewidth]{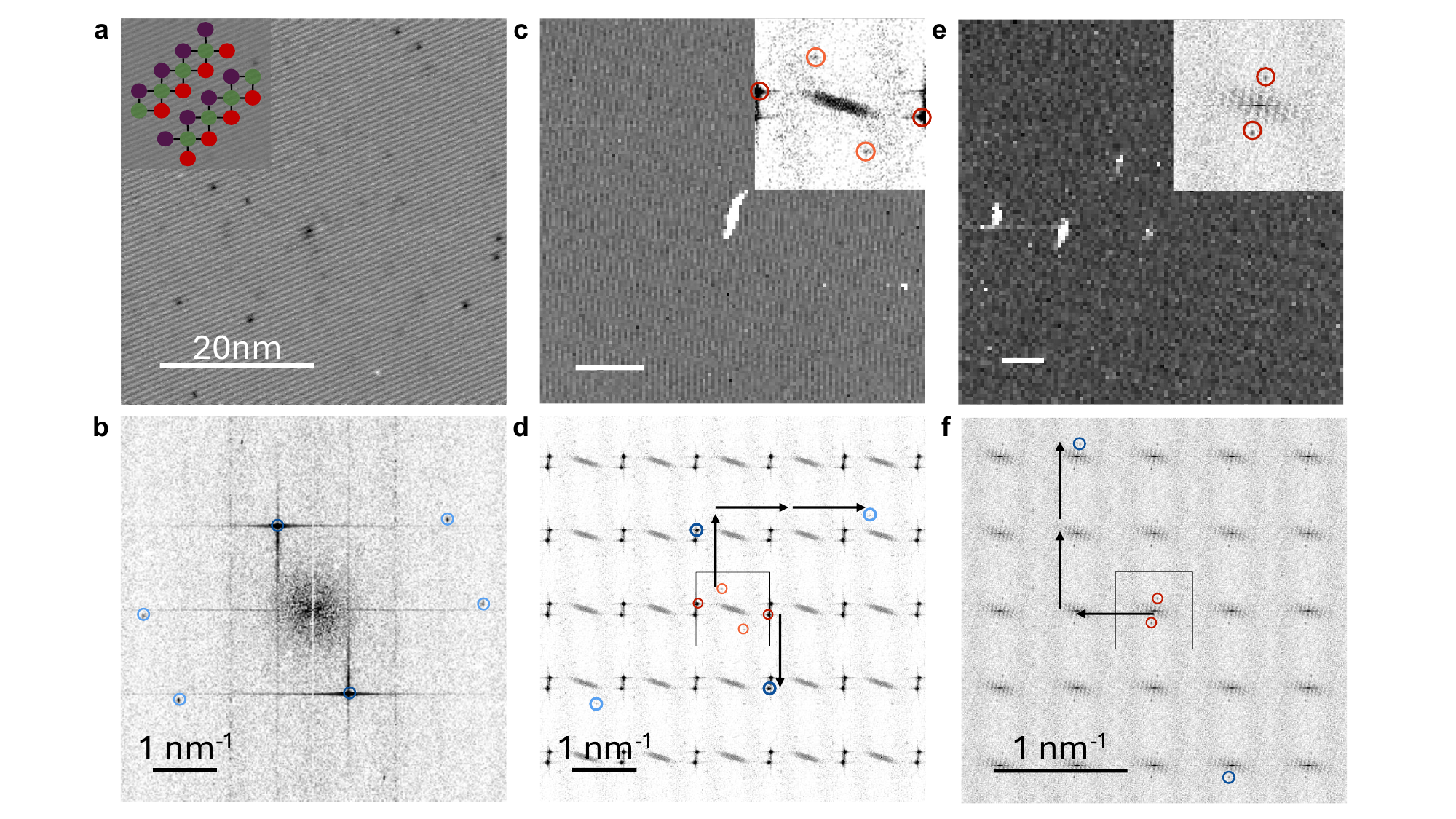}
    \caption{{\bf Replicas in STM topographic maps in UTe$_2$ of different sizes.} (a) STM topography map with atomic resolution. The atomic structure viewed from the surface plane is shown by disks in the upper left. Green is for U, red for one Te site and violet for the other Te site. The Fourier transform is shown in (b). We mark in dark blue peaks correspond to the Te chains and in light blue the other two relevant Bragg peaks of the surface atomic structure (the (011) plane, see Refs.\,\cite{talaveraCDW2025,Aishwarya2023,Gu2023,Aishwarya2024,LaFleur2024,science.adk7219}). (c) Larger scale topography. White scale bar has the same size as in (a). In the upper right inset we show the Fourier transform. Red circles show the replica peaks. (d) The same Fourier transform is copied to reach the size of (b) in reciprocal space. Blue circles are the peaks found in (b) and red circles those found from the Fourier transform shown in the upper right inset of (c). Black arrows connect blue and red circles. In (e) we show an slightly larger topography on a different field of view (white scale bar has the same size as in (a)). The Fourier transform is shown in the upper right inset is copied in (f) to reach approximately half the size of (b). Black arrows join red (replica Bragg peaks) with blue (atomic scale Bragg peaks) circles. The topographies (a,c,e) have been taken at $\mathrm{V_{Bias} = -50~mV, I_{Set} = 200~pA, T = 4.7~K}$. The black squares in (d,f) provide the boundary of the Fourier transforms of (c,e).}
    \label{fig:UTe2}
\end{figure}

Let us start with the example of topographic STM measurements of the atomic lattice of UTe$_2$. As we show in Fig.\,\ref{fig:UTe2}(a), a map of about 50\,nm in lateral size, atomic scale topography STM maps show a row-like pattern which is due to Te chains exposed after in-situ cryogenic cleaving. This image is similar to the images obtained in previous work and shows the (011) surface of the orthorhombic lattice of UTe$_2$\,\cite{talaveraCDW2025,Aishwarya2023,Gu2023,Aishwarya2024,LaFleur2024,science.adk7219}. The Te atoms at the surface present a distorted hexagonal ordering which is well identified in the Fourier transform by two main wave vectors, marked in light and dark blue in Fig.\,\ref{fig:UTe2}(b). The pattern is more intricate than a simple chain, and the two wave vectors are explained by the properties of the surface crystal structure\,\cite{talaveraCDW2025,Aishwarya2023,Gu2023,Aishwarya2024,LaFleur2024,science.adk7219}. Furthermore, there are large size atomically flat terraces. As we will see now, we can identify the Te chains in much larger images, made with the same number of points.

We show in Fig.\,\ref{fig:UTe2}(c) a map with about 100 nm lateral size. There are features in the map which provide peaks in the Fourier transform (upper right inset of Fig.\,\ref{fig:UTe2}(c)). To identify these peaks, and following the discussion above, we copy the Fourier transform as many times as needed to obtain a reciprocal space map of similar size as in the map where we observe the atomic rows (Fig.\,\ref{fig:UTe2}(b)). We find the pattern shown in Fig.\,\ref{fig:UTe2}(d). Using the vectors shown by black arrows, we can up-fold the modulations observed in Fig.\,\ref{fig:UTe2}(c) (the replica Bragg peaks are marked by light and dark red circles in Fig.\,\ref{fig:UTe2}(c,d)) to the atomic Bragg peaks (marked by light and dark blue circles in Fig.\,\ref{fig:UTe2}(d)). We find for the vectors highlighted in Fig.\,\ref{fig:UTe2}(d) $\mathbf{k}_{UTe_2,1}=(0,k_S)+\mathbf{k}_{replica,UTe_2,1}$ and $\mathbf{k}_{UTe_2,2}=(0,k_S)+(2k_S,0)+\mathbf{k}_{replica,UTe_2,2}$.

In an even larger map, Fig.\,\ref{fig:UTe2}(e) of nearly 200\,nm size, we identify one clear peak in the Fourier transform, that corresponds to the modulation in dark blue in Fig\,\ref{fig:UTe2}(b). By repeating the steps discussed above, we obtain Fig.\,\ref{fig:UTe2}(f), and can connect the replica Bragg peaks observed in Fig.\,\ref{fig:UTe2}(e) to the original atomic size modulation by adding $(2k_{S},0)$ and $(0,k_{S})$ to the replica. This shows that the Te chains occur with the same periodicity over close to micron length scales.

We note that reciprocal space provides a strong signal at small wavevectors. This consists of an ellongated ellipse in the inset of Fig.\,\ref{fig:UTe2}(c) and of the stripes shown in the inset of Fig.\,\ref{fig:UTe2}(e). As we show in the Appendix (Fig.\,\ref{SymDefect}(d)), this is due to the presence of defects producing additional interference patterns. For example, the white spots in Fig.\,\ref{fig:UTe2}(e) provide a small network of defects that lead to the stripes shown in the Fourier transform (inset of Fig.\,\ref{fig:UTe2}(e)).

\subsection{Modulations in the atomic lattice and in the superconducting density of states of FeSe}

\subsubsection{Atomic scale topography of FeSe.}

\begin{figure}[!htbp]
    \centering
    \includegraphics[width=1\linewidth]{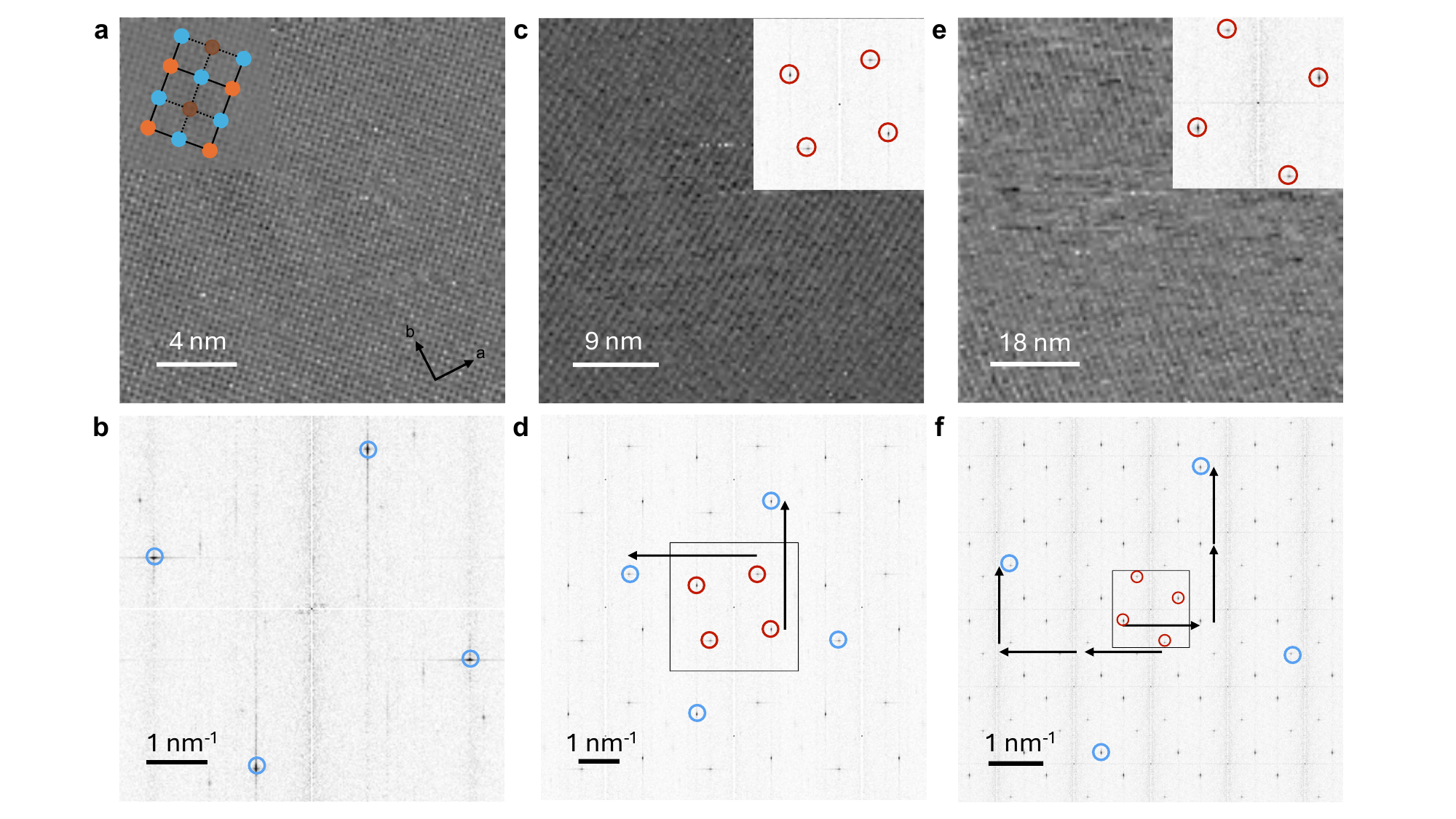}
    \caption{{\bf Replicas in STM topographic maps in FeSe of different sizes.} In (a) we provide a STM topography map of FeSe. In the upper left inset we show schematically the lattice of FeSe viewed from the surface (blue disks are Fe, orange and brown disks are the two relevant Se sites, see for more details Fig.\,\ref{SymmetriesFeSe}(a)). The crystalline lattice is shown by black arrows. In (b) we provide the Fourier transform and highlight with blue circles the position of the Bragg peaks of the atomic Se lattice. In (c) we provide a larger scale topography, and in the upper right inset we show the Fourier transform, marking with red circles the replica Bragg peaks. In (f) we show the Fourier transform of (c) copied several times. In (e) we show an even larger image, with the Fourier transform in the upper right inset. The replicated Fourier transform is shown in (f). Black arrows in (d,f) join replica Bragg peaks (red circles) with the atomic size modulation (blue circles). STM topographic maps in (a,c,e) are taken with $\mathrm{V_{Bias} = 6~mV, I_{Set} = 4~nA, T = 1~K}$, all with the same number of points. White lines in (a,c,e) and black lines in (b,d,f) provide scale bars in each panel. Black arrows join replica (red circles) with atomic modulation (blue circles) Bragg peaks.}
    \label{fig:FeSe}
\end{figure}

Now we turn to the layered superconductor FeSe. As shown by previous works, the square Se atomic lattice is usually visualized in STM topography (Fig.\,\ref{fig:FeSe}). In Fig.\,\ref{fig:FeSe}(a) we show an image with atomic resolution presenting the square Se lattice, and in Fig.\,\ref{fig:FeSe}(b) we present its Fourier transform. We now enlarge the field of view in the same area, maintaining the number of points and obtain Figs.\,\ref{fig:FeSe}(c,e). The latter figures also present a periodic signal, with wave vectors that do not correspond to the square Se lattice and are instead replicas of the atomic size modulation. By making the construction described above, we connect the replica Bragg peak positions with the atomic Bragg peaks by adding $(k_S,0)$ and $(0,k_S)$ to the replicas in Figs.\,\ref{fig:FeSe}(c,d) and $(2k_S,k_S)$ and $(k_S,2k_S)$ in Figs.\,\ref{fig:FeSe}(e,f). We note that the association is unambiguous. Of course, prior knowledge of the position and size of the atomic wave vectors is required. However, the modulation shown in Figs.\,\ref{fig:FeSe}(c,e) is unambiguously due to the periodic atomic lattice, even if the image is in itself under sampled.

Contrary to the case of UTe$_2$, here we do not find large size defects. The reciprocal space maps are free of features at small wavevectors. Furthermore, Bragg peaks remain sharp in Figs.\,\ref{fig:FeSe}(d,f) suggesting a homogeneous atomic lattice over the whole field of view.

We note that the square in Fig.\,\ref{fig:FeSe}(e) is distorted. This is not a measurement artifact, as can be seen from the construction in Fig.\,\ref{fig:FeSe}(f), which connects the distorted pattern marked with red circles with the undistorted pattern marked with blue circles. Simply, it is a consequence of the specific form of replicas (black arrows in Fig.\,\ref{fig:FeSe}(f)) used to connect atomic scale and large scale Bragg peaks. The replica signal is obtained by translation. If the translation requires a different number of wavevectors along the two main directions, $n_{x,y}$ with $n_{x} \neq n_{y}$, the replica Bragg peaks do not respect the symmetry of the atomic lattice. We can see this also nicely in Figs.\,\ref{fig:UTe2}(d,f), where the replica Bragg peaks are oriented along different directions than the atomic scale Bragg peaks.

\subsubsection{Atomic size modulation of the superconducting gap in FeSe.}

Having shown that atomic size topographic features are visible as replicas in large size images, we now concentrate on maps of the tunneling conductance. In FeSe numerous previous STM works have unveiled and analyzed spatially resolved superconducting features, including nematicity, behavior at nematic phase boundaries, vortex lattices, or, most recently, a novel pair density modulation when the inversion symmetry is broken at the surface by an enlarged c-axis constant\,\cite{science.aal1575,Bu2021,sciadv.aar6419,PhysRevX.5.031022,PhysRevLett.127.257001,Hoffman_2011,Kong2025,papaj2025pairdensitymodulationglide,zhang2024visualizinguniformlatticescalepair,Wei_2025,chen2025landautheorypairdensity}. FeSe has a tetragonal crystal structure which becomes slightly orthorhombic below 90\,K\,\cite{PhysRevB.96.024511,Baek2015,PhysRevLett.103.057002}. FeSe becomes superconducting below approximately 9\,K\,\cite{PhysRevB.96.024511,Bohmer_2018}. In-situ exfoliation reveals the van-der-Waals stacking along the c-axis and provides square lattices formed by Se atoms on the surface. The orthorhombicity is less than 0.5\% of the in-plane crystalline lattice and essentially indistinguishable using STM. The square Se atoms at the surface are separated by $\approx$ 3.8\,\AA.

\begin{figure}[!htbp]
\centering
\includegraphics[width=0.9\textwidth]{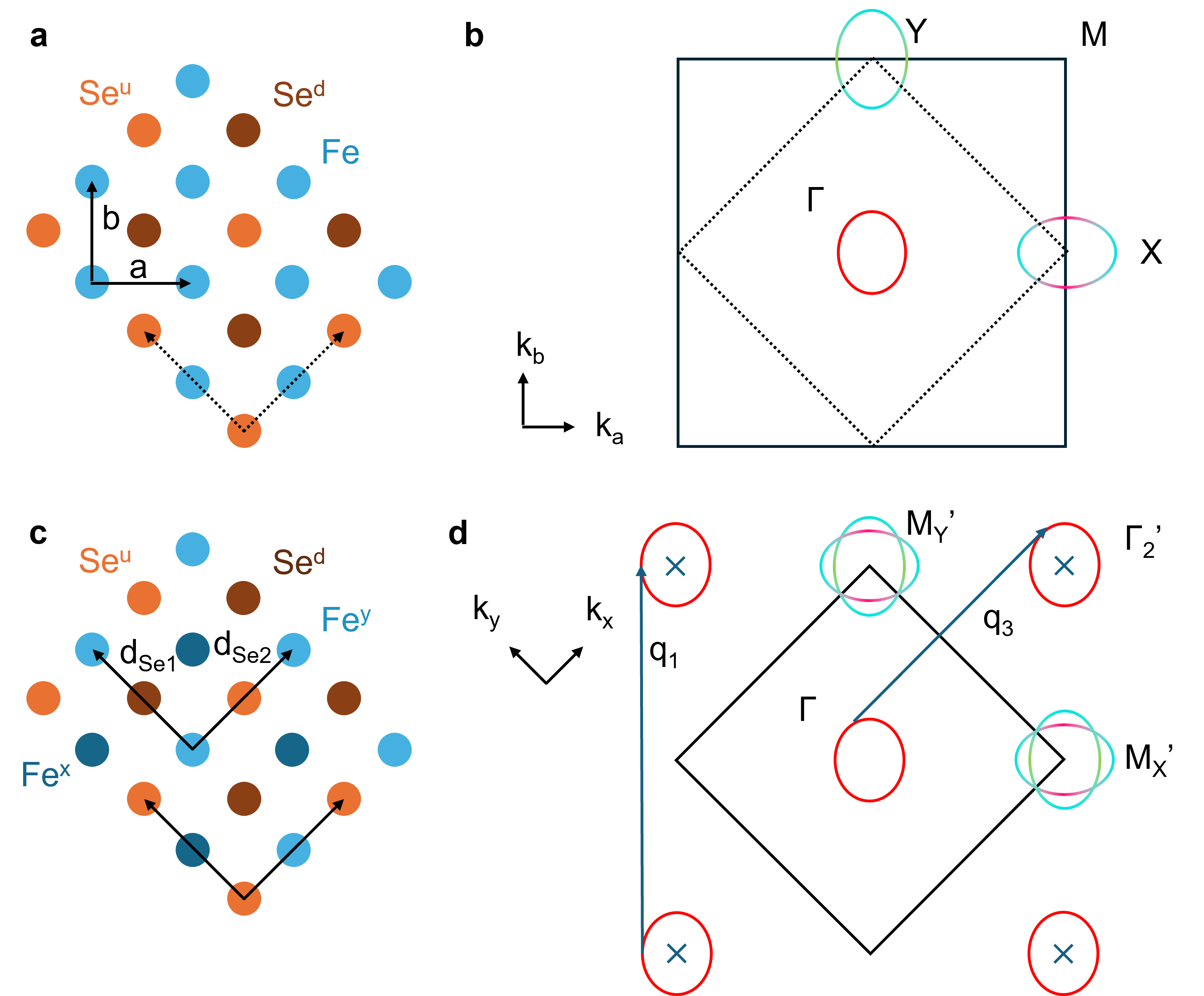}
\caption{{\bf Schematic representation of Fermi surface in FeSe with and without glide plane symmetry.} In (a) we show as orange and brown disks the Se lattice, with upper labels $^u$ and $^d$ to indicate the upper and lower Se planes, with respect to the Fe lattice (blue disks). In-plane crystalline axes are shown as black arrows. (b) Cartoon representation of the Fermi surface of the nematic phase of FeSe in reciprocal space. We show as a black square the Brillouin zone of the Fe lattice and as a black dashed line square the Brillouin zone of the Se lattice (the two-Fe lattice unit cell, also represented in real space by dashed arrows in (a)). Fermi surface pockets are shown as colored circles. Color corresponds to the orbital character, red for $d_{xz}$, green for $d_{yz}$ and blue for  $d_{xy}$. (c) FeSe lattice in absence of glide symmetry. The two Fe atoms (dark and light blue disks) are not equivalent. (d) Black square is the two-Fe unit cell. The Fermi surface pockets are now folded. Arrows represent wave vectors $q_1$ and $q_3$ which join the central Fermi surface hole pockets into each other. See Refs.\,\cite{Kong2025,papaj2025pairdensitymodulationglide} for more details.}\label{SymmetriesFeSe}
\end{figure}

The pair density wave modulation has been widely discussed and it is useful to briefly review a few aspects. The pair density modulation consists of a modulation of the superconducting gap at atomic scale originated in FeSe by a combination of nematicity, glide symmetry breaking and an anisotropic superconducting gap\,\cite{Kong2025,papaj2025pairdensitymodulationglide,zhang2024visualizinguniformlatticescalepair,Wei_2025,chen2025landautheorypairdensity}. The pair density modulation has been observed in very thin flakes of FeSe by imaging areas which are several nm of lateral size\,\cite{Kong2025,zhang2024visualizinguniformlatticescalepair,Wei_2025}. We present a simplified picture of the atomic and electronic structure of the surface of FeSe in Fig.\,\ref{SymmetriesFeSe}. More details are provided in Refs.\,\cite{Kong2025,papaj2025pairdensitymodulationglide,zhang2024visualizinguniformlatticescalepair,Wei_2025,chen2025landautheorypairdensity,Fernandes_2017,sym12091402}. The atomic structure viewed from the surface of FeSe is shown schematically in Fig.\,\ref{SymmetriesFeSe}(a), following \,\cite{Kong2025}. The upper Se atoms (orange disks in Fig.\,\ref{SymmetriesFeSe}(a)) are mostly visible in STM topographies. As mentioned above, the orthorhombicity is very small and the nematicity is essentially an electronic property which is seen as an in-plane anisotropic Fermi surface, shown in a simplified form in Fig.\,\ref{SymmetriesFeSe}(b). Experimentally, the distinguishing feature of the areas where a pair density modulation is observed is an enlarged c-axis constant, obtained when measuring the distance between layers on steps at the surface and observing distances approximately 20\% larger than $\delta c\approx$ 5.5\,\AA \,\cite{Kong2025}.  This implies that Fe atoms are located at slightly different distances from the Se planes, providing the structure shown in Fig.\,\ref{SymmetriesFeSe}(c). This breaks the glide symmetry of the FeSe blocks. As a consequence, the Fermi surface shows the structure presented in a simplified form in Fig.\,\ref{SymmetriesFeSe}(d), which has two wave vectors joining the central hole pockets. According to Refs.\,\cite{Kong2025,papaj2025pairdensitymodulationglide}, this provides a spatial modulation of the amplitude of the superconducting gap at atomic scale.

\begin{figure}[!htbp]
\centering
\includegraphics[width=0.9\textwidth]{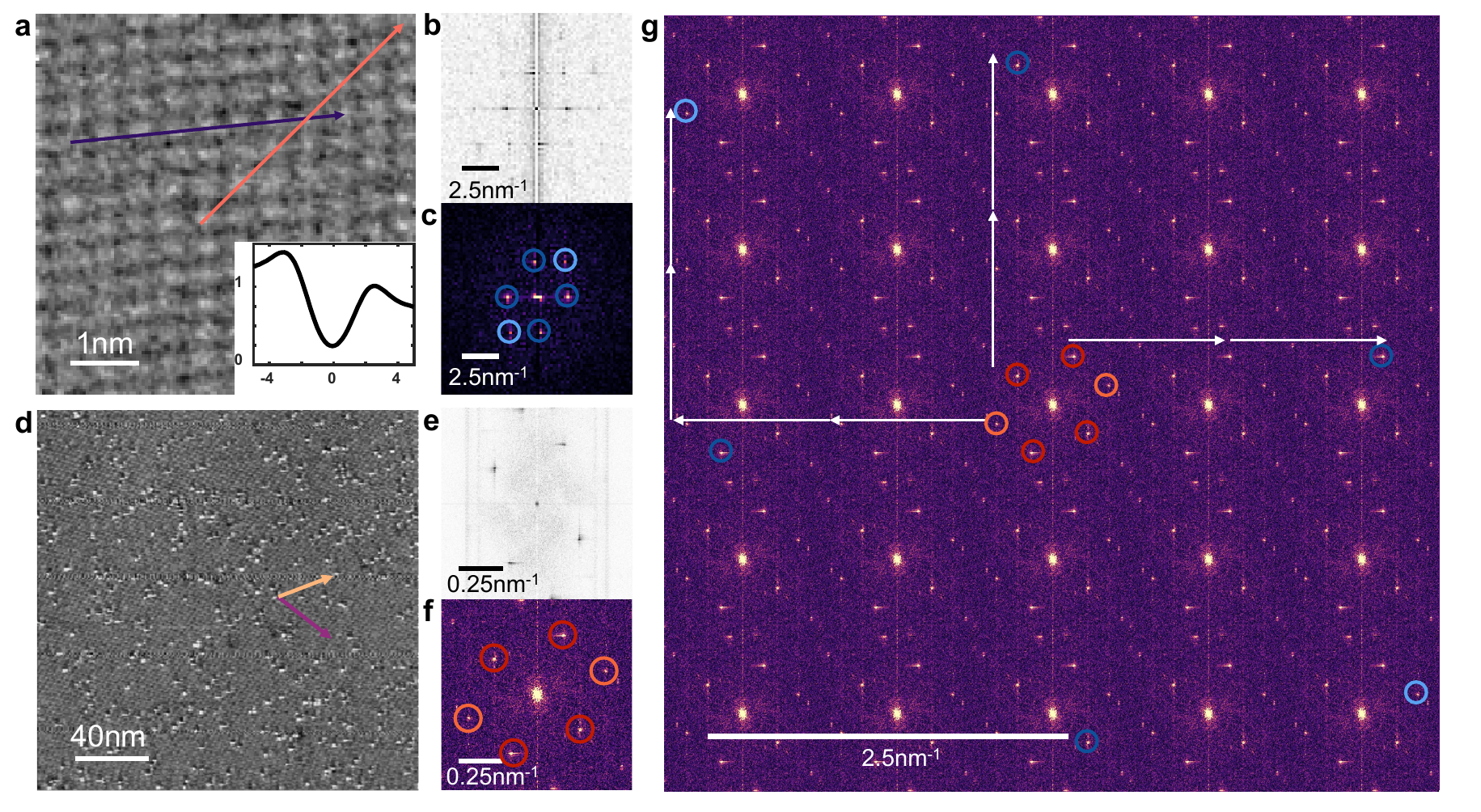}
\caption{{\bf Atomic scale gap size modulations in FeSe.} (a) Atomically resolved STM topographic image in FeSe taken at 4.2K and zero field (bias voltage of 10\,mV and current of 1.2\,nA). White scale bar is 1nm long. Inset shows the average tunneling conductance on the field of view. (b) Fourier transform of (a). In (c) we show the Fourier transform of the map of the size of the superconducting gap in the same field of view. Dark and light blue circles indicate Bragg peaks of the Fe and the Se atomic scale modulation, respectively. (d) STM topographic image taken in FeSe at a temperature of 1\,K (bias voltage of 6\,mV and tunneling current of 3\,nA). White scale bar is 40\,nm large. (e) Fourier transform of (d). (f) Fourier transform of the map of the superconducting gap acquired in the same field of view as (d). (g) We copy the Fourier transform (f) and show with dark blue and red circles the positions the atomic scale and replica Bragg peaks. White arrows connect replica with atomic scale Bragg peaks. Arrows in (a,d) provide the places where we took the profiles shown in Fig.\,\ref{PDMProfiles}.}\label{PDMFeSe}
\end{figure}

To present and discuss atomic size modulations of the superconducting density of states, we have made numerous tunneling conductance maps at different bias voltages. We focus on fields of view where we observe steps close to the atomically flat areas which have heights larger than $\delta c\approx$ 5.5\,\AA\, occasionally reaching $\delta c\approx$ 7\,\AA. These steps occur occasionally due to the in-situ mechanical exfoliation procedure, as shown for instance in Refs.\,\cite{PhysRevB.87.094502,PhysRevLett.102.176804}. In these areas, we find atomic scale changes in the superconducting tunneling conductance (Fig.\,\ref{PDMFeSe}). In images a few nm in lateral size we find results resembling those found in\,\cite{Kong2025,papaj2025pairdensitymodulationglide,zhang2024visualizinguniformlatticescalepair,Wei_2025}.  Although a precise identification of a pair density wave requires more studies, for example including the Josephson effect\cite{Escribano2025}, we will discuss results obtained from measurements of the tunneling conductance.

We show the atomic lattice in Fig.\,\ref{PDMFeSe}(a) and the spatially averaged superconducting gap in the lower inset of Fig.\,\ref{PDMFeSe}(a). We measure the superconducting gap at each position by measuring the position in bias voltage of the quasiparticle peaks for both negative and positive bias voltages and taking their average. We make then real space maps of this quantity, which look similar to the atomic lattice pattern shown in Fig.\,\ref{PDMFeSe}(a). We then make the Fourier transform and identify, in the gap map, the main atomic scale wave vectors, marked with dark and light blue circles in Fig.\,\ref{PDMFeSe}(c).

To address whether such a modulation of the superconducting density of states is observed over large length scales, we now use R-STM. We acquire an image of over 200 nm in lateral size with just 256 $\times$ 256 points. We show the corresponding topography in Fig.\,\ref{PDMFeSe}(d) and its Fourier transform in Fig.\,\ref{PDMFeSe}(e). We show the Fourier transform of the map of the superconducting gap (obtained again by measuring the position of the quasiparticle peaks in the bias voltage, averaged over negative and positive bias) in Fig.\,\ref{PDMFeSe}(f). We connect replicas to the atomic size modulation in Fig.\,\ref{PDMFeSe}(g). This suggests that the atomic scale gap modulation persist over large length scales.

\begin{figure}[!htbp]
\centering
\includegraphics[width=0.9\textwidth]{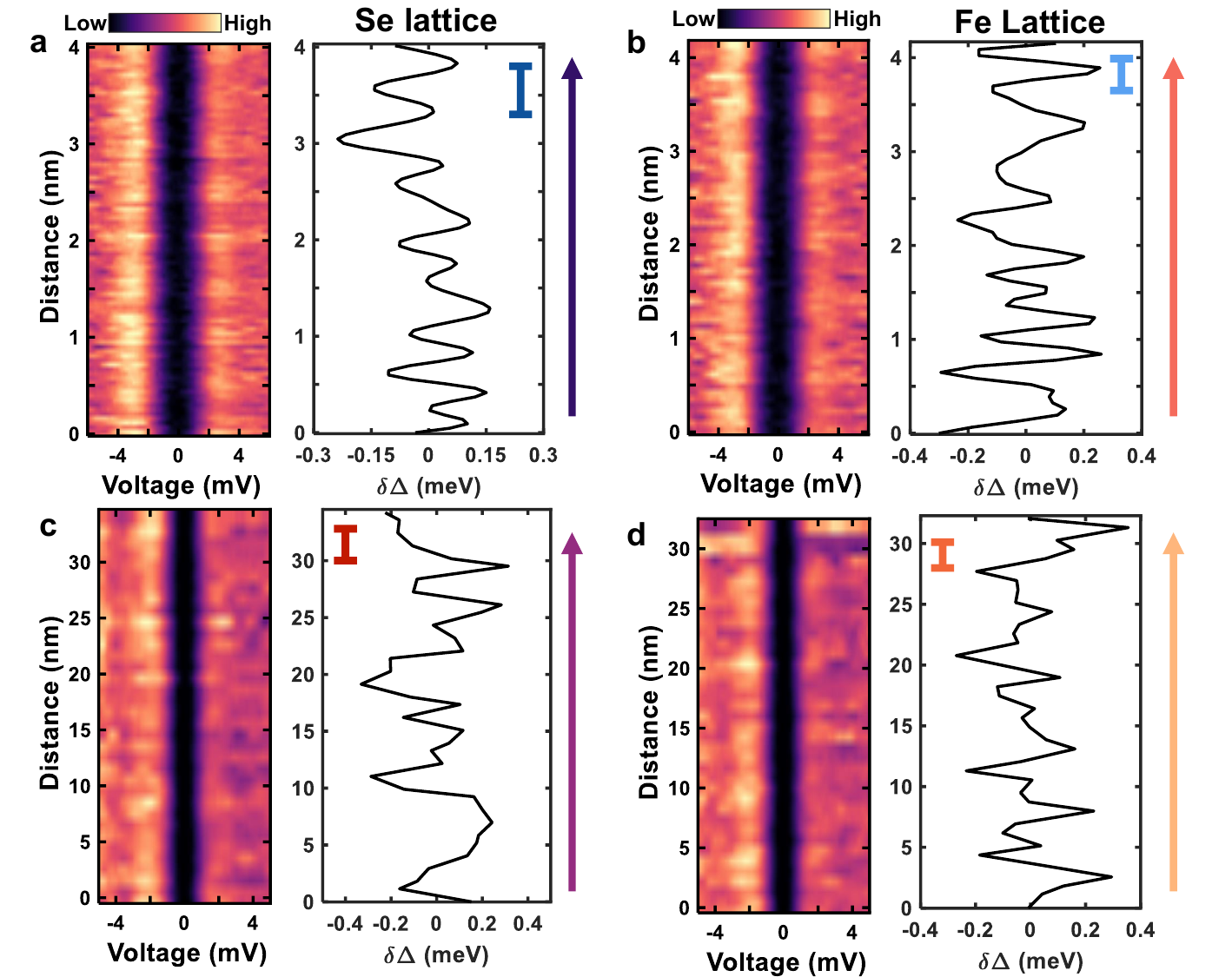}
\caption{{\bf Line profiles of the superconducting gap; atomic vs replica scales.} (a) In the left panel we show with a color scale the tunneling conductance versus bias voltage along the dark violet line shown in Fig.\,\ref{PDMFeSe}(a). In the right panel we show the spatial variation of the magnitude of the superconducting gap $\delta \Delta$, obtained from the quasiparticle peak position (averaging positive and negative bias voltages). (b) Same along the dark orange arrow shown in Fig.\,\ref{PDMFeSe}(a). (c,d) Same, but along the light violet and light orange arrows of  Fig.\,\ref{PDMFeSe}(d). Notice the change in the length scale when comparing (a,b) with (c,d). We show as colored vertical bars with an `I' shape the wavelengths corresponding to the Bragg peaks of the atomic scale modulations. Dark blue in (a) for the dark blue atomic scale Bragg peaks in Fig.\,\ref{PDMFeSe}(c), light blue in (b) for the light blue atomic scale Bragg peaks in Fig.\,\ref{PDMFeSe}(c), red in (c) for the red replica Bragg peaks in Fig.\,\ref{PDMFeSe}(d) and orange in (d) for the orange replica Bragg peaks in Fig.\,\ref{PDMFeSe}(d).}\label{PDMProfiles}
\end{figure}

To discuss this in more detail we show in Fig.\,\ref{PDMProfiles}(a-d) profiles of the tunneling conductance as a function of the distance. We show the calculated relative variation of the gap size ($\delta \Delta$). We show data along the Se lattice (Fig.\,\ref{PDMProfiles}(a,c)) and along the Fe lattice (Fig.\,\ref{PDMProfiles}(b,d)). The corresponding gap modulations have a completely different wavevector when measured with atomic precision (Fig.\,\ref{PDMProfiles}(a,b)) as compared to measurements using R-STM (Fig.\,\ref{PDMProfiles}(c,d)) and are connected through the replica construction shown by the white vectors in Fig.\,\ref{PDMFeSe}(g). The vertical `I'-shaped bars in Fig.\,\ref{PDMProfiles}(a-d) provide the real space distance associated with the corresponding Bragg peaks shown in Fig.\,\ref{PDMFeSe}(c,f).

Thus, we see that R-STM bridges the distance between atomic size and near to micron sized modulations, and suggests that gap modulations might occur coherently over large length scales. We remark that we have significant low wavevector signal in the Fourier transform of the large scale maps (Fig.\,\ref{PDMFeSe}(f)). This suggests that there are many defects perturbing superconductivity over the image. Furthermore, Bragg peaks are broadened in the Fourier transform of the large scale maps (Fig.\,\ref{PDMFeSe}(f)) suggesting that there are domains over such large length scales.

\section{Discussion}\label{sec4}

We have shown for two different examples that R-STM is a powerful new technique to study atomic size modulations over very large length scales. The modulations can be easily identified as replica Bragg peaks in the Fourier transform of the maps.

It is useful to compare our technique with sparse sampling\cite{nakanishi2016,Oppliger2020,Zengin2021,Oppliger2022}. Both techniques aim to reduce the acquisition time of STM data, but take very different paths to achieve that. Sparse sampling is a very flexible and powerful technique for exploring large areas, and can work generally for any kind of dataset. However, the system under study has to be limited to a few narrow features in order to achieve significant time savings.
On the other hand, replicas don't require the data to be sparse, but rather than it is localized as much as possible in reciprocal space, so that aliasing does not map different features into the same wavevectors. This means that for different kinds of datasets, one technique will perform better than the other. One significant benefit of sparse sampling is the ability to use informed sampling, tailoring the sampling  pattern to the area of interest to optimize the signal. A relevant issue in sparse sampling is that it involves additional computations, both before data acquisition to calculate an optimized path for the tip, and after data acquisition\cite{nakanishi2016,Oppliger2020,Zengin2021,Oppliger2022}. R-STM is directly implemented in any scanning probe set-up, and even used a post-processing technique on existing data. The best results will be achieved for periodic signals and by carefully choosing the sampling frequency for the given data.

The observation of coherent atomic size images at the surface over large length scales, demonstrated here in Figs.\,\ref{fig:UTe2},\ref{fig:FeSe}, is certainly an insight which was, to the best of our knowledge, never obtained before using STM. However, it is somehow to be expected, because X-ray scattering of the same samples provides single crystalline Bragg peaks, implying a coherent atomic lattice in the whole sample. While atomic size impurities are identified in the Fourier transform of the large scale images, the absence of large surface steps over such length scales shows that there are very well-defined cleaving planes \,\cite{Herrera2021,Herrera2023}. But, other than that, our result shows that the surface is as crystalline as the bulk.

However, large scale atomic size modulations of the superconducting gap with the same wavelength could not have been anticipated before. Our results suggest that atomic size modulations of the superconducting density of states can be eventually identified over large length scales using R-STM. This supports the idea that the occurrence of superconducting states with a finite momentum, as the pair density wave state, could possibly be generic long-range feature of the superconducting state\cite{AgterbergDavis2020,Du2020,Wang2021,Hamidian2016,Chen2022,Zhao2023, Liu2023,Chen2021,Deng2024,Gao2024,Sumita2025,Kong2025,Gu2023,PhysRevB.77.134505}.

Prior experiments often relate the pair density and charge density waves\cite{AgterbergDavis2020,Du2020,Wang2021,Hamidian2016,Chen2022,Zhao2023, Liu2023,Chen2021,Deng2024,Gao2024,Sumita2025,Kong2025,Gu2023,PhysRevB.77.134505}. The latter are sometimes observed using other techniques, as X-ray and other macroscopic measurements as for instance transport. The atomic size gap modulations of the superconducting density of states can only be studied until now using STM. It would therefore be very interesting to apply R-STM to cuprates, kagome superconductors, chalcogenides, heavy fermions and other systems where finite q superconducting states are suspected or have been found using local probes\cite{AgterbergDavis2020}.
 
\section{Conclusion}

In conclusion, we have shown that large scale images with R-STM provide accurate information about atomic modulations at long length scales. The unequivocal association of replicas from atomic modulations can be used to demonstrate coexisting phenomena at different length scales. We can think of identifying and studying atomic density modulations and the superconducting vortex lattice at the same time, unveiling the spatial dependencies for different energies close to the Fermi level. For instance, sometimes vortex lattice images present periodic stripes at length scales which are much larger than atomic distances (for example Figure 3a in Ref.\,\cite{bisset2025determiningsuperconductingorderparameter} or Figure 1 of Ref.\,\cite{Sharma2025}). It would be interesting to analyze if these stripes are due to replicas of smaller sized atomic size modulations. R-STM will be also very helpful to associate atomic size features in semimetals and semiconductors, as reconstructions or local charge density modulations around atomic size impurities, with large scale charge density modulations\cite{PhysRevB.97.014505,Yay24}.

Studying large areas with R-STM allows us to simultaneously visualize the atomic lattice and other atomic scale modulations, such as superconducting gap variations or moire superlattices, provided that the latter are known from small scale atomic studies. However, measurements on large areas have a reduced sensitivity to structural features such as defects and domain boundaries (see Appendix). A relevant source of uncertainty in STM could occur in compounds where STM maps show strong features coming from different layers, not just the outer or immediate next layer. This is not very common because the largest contributions often come from the very last layer and the latter is often formed by a single atomic species. However, connecting atomic scale with replica Bragg peaks can be very difficult for complex atomic scale arrangements at the surface.

R-STM can be directly applied to any local probe technique, bridging length scales which are separated by up to about two orders of magnitude as long as the relevant phenomena are characterized by small scale periodic modulations. The technique requires that there is no additional broadening induced by the simple fact of measuring large scale distances. This is not the case in STM, because tunneling occurs at very confined length scales. But it could be important for other scanning probe techniques. If the local nature of the probe is maintained, R-SPM (Scanning Probe Microscopy) could become very interesting for other probes. For example, magnetic patterns in scanning magnetic techniques\cite{adma202307195,Auerbach2025,PhysRevX.15.021041}, electronic patterns with multiple periodicities combined with disorder\cite{PhysRevResearch.3.013022}, atomic size features in nanostructured samples\cite{Phark2025}, periodic features observed by atomic force microscopy\cite{RevModPhys.75.949}, or heat dissipation patterns probed by local thermal techniques\cite{Halbertal2016,BermudezPerez2024}.

\section{Appendix}

\subsection{Domains}

\begin{figure}[!htbp]
\centering
\includegraphics[width=0.8\textwidth]{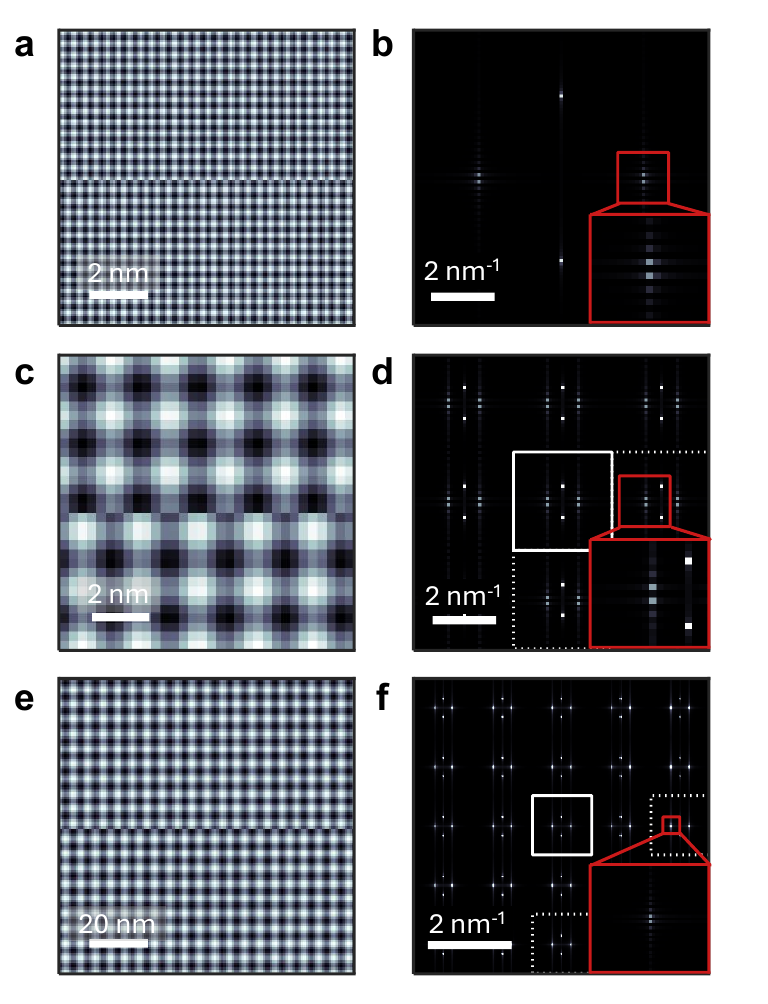}
\caption{{\bf Influence in the replica Bragg peaks of shifts in the modulation.} In (a) we show in a color scale a sinusoidal square lattice with a $\pi$ shift at the center of the image. In the lower bottom inset of (b) we zoom into the Bragg peak split by the additional long wavelength modulation. In (c) we show the same field of view as (a) but with a smaller number of points. We show the corresponding Fourier transform in (d), within the white square. The Fourier transform has been copied to reach a Fourier space map of the same size as the one in (b). In (e) we show the same pattern as in (a), with the same number of points but on a larger field of view and in (f) the Fourier transform (within white square), again copied to reach the same size in Fourier space as (b). corresponding replicated Fourier map is shown in (f). Red squares indicate zooms around the split peaks in Fourier space.}\label{SymDomains}
\end{figure}

To discuss the role of domains in the replica signal we have simulated a square lattice including a $\pi$ shift at the $y=0$ line (Fig.\,\ref{SymDomains}). We take a signal of the form $sin(\vec{\mathbf{q}}\cdot\vec{\mathbf{r}}+\phi(\vec{r}))$, and we add a $\pi$ step in $\phi(\vec{r})$ in the middle of the map along the y direction. The shift simulates a domain in a charge density wave or a step in a crystal lattice. The shift introduces a new long wavelength periodicity (half the image size for a single phase slip as shown in Fig.\,\ref{SymDomains}(a)). This leads to Bragg peak splitting along the y-axis of reciprocal space (Fig.\,\ref{SymDomains}(b)) and a reduction of the amplitude, as compared to the unsplit peaks. We show the result of undersampling in the same field of view as Fig.\,\ref{SymDomains}(a) in Fourier space in Fig.\,\ref{SymDomains}(d,f). The peak splitting is carried over into the replica Bragg peaks. For an increased field of view, Fig.\,\ref{SymDomains}(e), the distance between the splitted replica Bragg peaks decreases (Fig.\,\ref{SymDomains}(f)). Thus, we conclude that replica Bragg peaks and the method described here remain for domains having different phases in modulated patterns. However, the signature for the presence of domains, which is essentially Bragg peak splitting, is more difficult to resolve in large size images. The decrease in the distance between split Bragg peaks can be more difficult to separate from other sources of Bragg peak broadening, discussed in the following.

\subsection{Noise}

\begin{figure}[!htbp]
\centering
\includegraphics[width=0.7\textwidth]{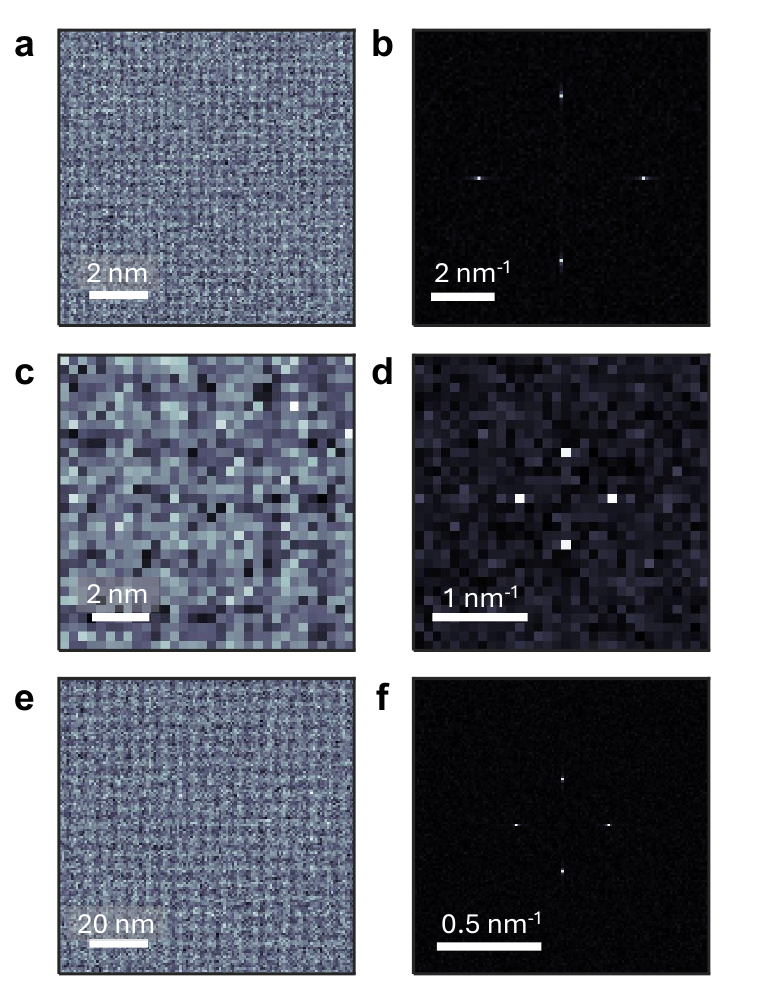}
\caption{{\bf Influence of random noise in the replica Bragg peaks.} In (a) we in a color scale a sinusoidal square lattice with added Gaussian noise and in (b) its Fourier transform. In (c), we show the same image as (a) but with a smaller number of points. In (d) we show the Fourier transform of (c). In (e) we show the same sinusoidal square lattice with noise, having the same number of points as (a) but on a larger field of view. The corresponding Fourier transform is shown in (f).}\label{SymNoise}
\end{figure}

To investigate the role of noise in the replica signal, we provide a periodic image with random disorder in Fig.\,\ref{SymNoise}(a). The noise is a Gaussian randomized background signal added to the periodic modulation. The Bragg peaks shown in Fig.\,\ref{SymNoise}(b) are slightly broadened due to noise and there is a white background in Fourier space (not shown in Fig.\,\ref{SymNoise}(b)). When undersampling, Fig.\,\ref{SymNoise}(c), the real space map becomes more flurry and it is difficult to distinguish a modulation. However, we find four clear replica peaks in reciprocal space (Fig.\,\ref{SymNoise}(c)). Increasing the field of view (Fig.\,\ref{SymNoise}(e)) provides again clear replica signals (Fig.\,\ref{SymNoise}(f)). The latter can be traced back to the original modulation shown in Fig.\,\ref{SymNoise}(a) by the method described here (for clarity, we do not provide copies of the replica signal in Fig.\,\ref{SymNoise}(d,f)). Thus, we conclude that replicas are in principle insensitive to noise, except that the signal to noise ratio of the Bragg peak is reduced.

\subsection{Defects}

\begin{figure}[!htbp]
\centering
\includegraphics[width=0.7\textwidth]{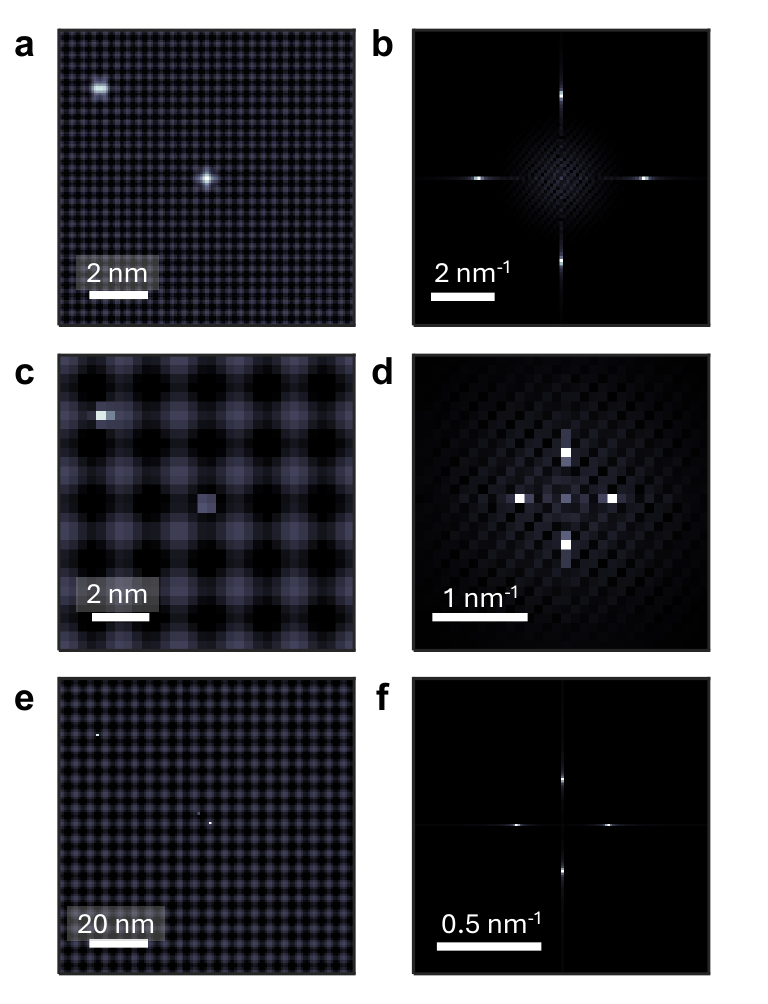}
\caption{{\bf Influence of point like and extended defects in the replica Bragg peaks.} In (a) we a sinusoidal square lattice including a brighter spot at the center and a defect on the top left corner. In (b) we show its Fourier transform. In (c) we show the same image as in (a) but with a smaller number of points. The Fourier transform of (c) is shown in (d). In (e) we show the same sinusoidal square lattice with the same number of points as in (a) but on a larger field of view. For clarity, we placed the two defects shown in (a) and (c) close to the center of (f). We also placed, in addition, a third defect close to the top left corner (white point at top left of (e)).}\label{SymDefect}
\end{figure}

In Fig.\,\ref{SymDefect}(a) we present a periodic signal containing two defects in Fig.\,\ref{SymDefect}(a). The defects are obtained by modifying the periodic function at the center of the image and on the upper left corner. The defects introduce stripes at small wavevectors in reciprocal space (Fig.\,\ref{SymDefect}(b)), because of the interference between the observed modulations and the defects. The wavelength of the stripes is related to the distance between defects. When we reduce the number of points (Fig.\,\ref{SymDefect}(c,d)), the Fourier transform provides the replica Bragg peaks (Fig.\,\ref{SymDefect}(d)) and replicas of the stripes. Increasing the size of the image (Fig.\,\ref{SymDefect}(e,f)) leads to a similar result.

In addition, the presence of defects introduces a new periodicity. Similarly as for domains, this results in modified Bragg peaks. For the patterns shown in Fig.\,\ref{SymDefect}(a,c,e), Bragg peaks are elongated along $x$ and $y$ axis in Fourier space Fig.\,\ref{SymDefect}(b,d,f).
 
\subsection{Drift}

We note that the atomic lattice and gap map at some locations can be slightly distorted (Fig.\,\ref{PDMFeSe}(a-c)). This might be due to drift or creep of the piezo or the voltage source and are an artifact of scanning. The replicas carry this distortion to large length scales. As long as the scanning speed (time required to move the tip between two points at the same distance) remains similar at small and large scales, we do not expect any influence of drift in the method discussed here. If the drift changes in different parts of the image, the  Bragg peaks will be blurred and increase their size, which will make it more difficult identifying individual Bragg peaks in replicas too.

\backmatter

\bmhead{Acknowledgements}

This work was supported by the Spanish Research State Agency (PID2020-114071RB-I00, PID2023-150148OB-I00, TED2021-130546B\-I00 and CEX2023001316-M), Comunidad de Madrid through project TEC-2024/TEC-380 “Mag4TIC” and PhD thesis support (PIPF-2023/TEC-30853 and PIPF-2023/TEC-30683), and the EU through grant agreement No 871106. We acknowledge the QUASURF project (SI4/PJI/2024-00199) funded by the Comunidad de Madrid through the agreement to promote and encourage research and technology transfer at the Universidad Aut\'onoma de Madrid. We acknowledge collaborations through EU program Cost CA21144 (superqumap.eu). We also acknowledge SEGAINVEX for support in design and for construction of the electronics of the STM.

\bibliography{biblio_R_STM}

\end{document}